\documentstyle[prb,aps,multicol]{revtex}

\begin{document}

\title{Density correlations in the half-filled Hubbard
       model}
   
\author{Fabian H. L. E\ss{}ler$^{(a)}$ and Holger Frahm$^{(b)}$}
\address{
$^{(a)}$Department of Physics, Theoretical Physics, Oxford University,
1 Keble Road, Oxford, OX1 3NP, United Kingdom\\
$^{(b)}$Institut f\"ur Theoretische Physik, Universit\"at Hannover,
         D-30167~Hannover, Germany
}
\date{March 1999}
\maketitle
\begin{abstract}
We consider density-density correlations in the one-dimensional
Hubbard model at half filling. On intuitive grounds one might expect
them to exhibit an exponential decay.  However, as has been noted
recently, this is not obvious from the Bethe Ansatz/conformal field
theory (BA/CFT) approach.  We show that by supplementing the BA/CFT
analysis with simple symmetry arguments one can easily prove that
correlations of the lattice density operators decay exponentially.
\end{abstract}
\pacs{71.10Fd 71.10.Pm 75.10.Jm}

\begin{multicols}{2}
Recently, density correlation functions have been studied for various
multicomponent one-dimensional electron systems within the framework
of the bosonization approach \cite{Emery99,Azariax}. An important
issue raised in these works is the question whether or not these
correlators exhibit a power-law behaviour at large distances in the
presence of a Mott-Hubbard gap (see also
Refs.~\onlinecite{Emery79,Voit88}).  For the particular case of the
half-filled, one-dimensional Hubbard model a bosonization analysis
yields the expected exponential decay, which at first sight appears to
be at odds with the results obtained from the Bethe Ansatz
\cite{frko:90} (see e.g. [20] of Ref.~\onlinecite{Emery99}).

In one spatial dimension, the asymptotic behaviour of correlation
functions can be analyzed by means of the BA/CFT approach. Furthermore,
all possible critical dimensions can be obtained from the exact solution
\cite{frko:90}. Since the expansion of the lattice operators in
terms of the conformal fields is not known explicitly, the resulting
expressions contain unknown amplitudes. In particular, the leading
term in the BA/CFT expression for the correlation function under
question could actually vanish. This is the origin of the controversy
mentioned above.  In certain cases, including the one discussed here,
it is possible to resolve this issue by employing symmetry
considerations of the underlying microscopic model. 

In this note, we are interested in connected correlation functions of
electron densities at half filling. Using the selection rules of
conservation of numbers of electrons with spin up and down, one can
show from the exact solution (specialized to the case of zero magnetic
field) that
\begin{eqnarray}
&&G_{nn}(t,x)= \langle (n(t,x)-\langle n \rangle)(n(0,0)-\langle n
\rangle)\rangle
\nonumber\\
&&\quad\longrightarrow
A_1\frac{\cos(\pi x/a_0)}{|x+ivt|}+
A_2\frac{x^2-v^2t^2}{(x^2+v^2t^2)^2}+\ldots\ ,
\label{hfb}
\end{eqnarray}
where $v$ is the spin velocity, $a_0$ is the lattice spacing and
$A_{1,2}$ are unspecified amplitudes that depend on the physical
parameters of the problem i.e. interaction strength and in the general
case magnetic field. Note that in (\ref{hfb}) we have omitted similar,
subdominant terms.

We will now show that by using the {\sl full} set of selection rules
for the half-filled Hubbard model, one can establish that the
amplitudes of all contributions to (\ref{hfb}), that decay
algebraically, are identically zero and that the correlator thus
decays exponentially.

The Hamiltonian of the half-filled Hubbard model is of the form
\begin{eqnarray}
 {\cal H}(U) &=& -\sum_{\langle ij\rangle} 
 \left\{ c^\dagger_{i\sigma}c_{j\sigma} + h.c.\right\}
\nonumber\\
&&
 + U \sum_i \left(n_{i\uparrow}-\frac{1}{2}\right)
	      \left(n_{i\downarrow}-\frac{1}{2}\right)
\label{hamil}
\end{eqnarray}
Here $c_{i\sigma}$, $c_{i\sigma}^\dagger$ are canonical annihilation
and creation operators for electrons with spin $\sigma$ on site $i$,
$n_i= \sum_\sigma n_{i\sigma} = \sum_\sigma c^\dagger_{i\sigma}
c_{i\sigma}$ is the number operator density for electrons.

In addition to the $SU(2)$ invariance in the spin degrees of freedom
the Hubbard Hamiltonian on a bi-partite lattice commutes with the
generators of a second $SU(2)$ related to particle-hole symmetry
\cite{HeLi71,yang:89}.  In
the one dimensional case the local generators of this $\eta$-pairing
$SU(2)$ are
\begin{eqnarray}
 & \eta_j^+ = (-1)^j c_{j\uparrow}c_{j\downarrow},\quad
  \eta_j^- = \left(\eta_j^+\right)^\dagger,&
\nonumber\\
 & \eta_j^z = {1\over2} \left(1-n_{j}\right),&
\label{etadef}
\end{eqnarray}
satisfying commutation relations $[\eta_j^\pm,\eta_j^z]
=\pm\eta_j^\pm$ and $[\eta_j^+,\eta_j^-]=2\eta_j^z$.  Combining the
two symmetries one obtains the well-kown $SU(2)\otimes
SU(2)/Z_2=SO(4)$-symmetry of the Hubbard model. 

At half filling the ground state $|\Omega\rangle$ is an
$SO(4)$-singlet \cite{lieb:89,esko:94l}.  Denoting by $n(t,x)$ the
density operator at $x=a_0 j$ and time $t$ we want to determine the
asymptotic behaviour for $j\to\infty$ of the connected correlation
function
\begin{eqnarray}
 G_{nn}(t,x)
 &=&4 \langle \eta^z(t,x) \eta^z(0,0)\rangle
\nonumber\\
 &=&4 \sum_n \langle\Omega| \eta^z(t,x)|n\rangle
  \langle n| \eta^z(0,0)|\Omega\rangle 
\nonumber\\
 &=&4\sum_n \exp\left({i(\epsilon_nt-k_nx)}\right)
 \left|  \langle n| \eta^z(0,0)|\Omega\rangle \right|^2
\label{corrnn}
\end{eqnarray}
Hence the matrix elements
\begin{equation}
  \langle n | \eta^z_1 |\Omega\rangle 
\label{matel}
\end{equation}
determine which intermediate states will contribute to the asymptotic
behaviour of (\ref{corrnn}).

In the BA/CFT approach to correlation functions in integrable models one
usually proceeds as follows: Firstly, one calculates the finite-size
scaling behaviour of the energies $\epsilon_n$ and momenta $k_n$ of
the low-lying states entering the expression (\ref{corrnn}).
Secondly, one uses the quantum numbers of the operator entering the
matrix element to eliminate ``incompatible'' intermediate states.
Finally, the CFT relation between finite-size energies and scaling
dimensions is used to obtain the asymptotic (power law) behaviour of
correlation functions by summing over the remaining low-lying
intermediate states.

In Ref.~\onlinecite{frko:90} only particle numbers for spin-$\sigma$
electrons were used in the second step.  This corresponds to taking
into account only quantum numbers associated with the Cartan
generators $S^z$ and $\eta^z$ but not those with the total spins
$\vec{S}^2$ and $\vec{\eta}^2$.  As a result one obtains (\ref{hfb}),
which seems to indicate that $G_{nn}$ decays algebraically for large
distances in spite of the presence of a charge gap at half filling.
We will now show, by considering the {\sl full} set of $SO(4)$ quantum
numbers, that only intermediate states with a gap contribute to
(\ref{matel}) and the density correlations decay exponentially.

Let us calculate the total $\eta$-spin quantum number of the state
$\eta^z_1 |\Omega\rangle$:  using the commutation relations between
the $\eta$-pairing operators (and the fact that $|\Omega\rangle$ is 
a singlet under the $\eta$-pairing SU(2)) one easily obtains
\begin{equation}
  \vec{\eta}^2 \left(\eta^z_1 |\Omega\rangle \right) =
  \left[ \vec{\eta}^2 ,\eta^z_1\right] |\Omega\rangle =
  2 \left(\eta^z_1 |\Omega\rangle \right),
\end{equation}
showing that this state is a triplet of the $\eta$-pairing $SU(2)$.
As a consequence, only intermediate states which are $\eta$-pairing
triplets can contribute to the correlation function (\ref{corrnn}).

As is well known only \emph{pure} spin excitations are gapless in the
repulsive half filled Hubbard model.  However, these are all singlets
of the $\eta$-pairing $SU(2)$ \cite{esko:94l}.  Therefore the
corresponding matrix elements (\ref{matel}) vanish identically and
cannot contribute in $G_{nn}$.  We conclude that the lowest energy
intermediate states with non-zero matrix elements (\ref{matel}) are
holon-antiholon scattering states with energy above the Mott-Hubbard
gap.  As a result the density-density correlation function exhibits
exponential decay at large distances for any positive $U$
\begin{equation}
  G_{nn}(t,x) \longrightarrow \exp(-\alpha x)\ ,\quad x\to\infty.
\end{equation}
The determination of $\alpha(U)$ is an interesting open problem. On
general grounds we expect it to be proportional to the holon gap,
which is given by\cite{lw,ovch:70} (see also Ref.~\onlinecite{Staff93})
\begin{eqnarray}
\Delta(U)&=&\frac{U}{2}-2+
  4\sum_{n=1}^\infty (-1)^n\left[\sqrt{1+(nU/2)^2}-nU/2\right]\
\nonumber\\
  &\sim& \frac{4}{\pi}\sqrt{U} \exp(-2\pi/U)\ ,
  \qquad {\rm for~}U\to0\ .
\end{eqnarray}

If one considers correlation functions of ``point-split''
densities such as $N^{\rm (ps)}_j=\sum_\sigma 
c^\dagger_{j\sigma}c_{j+1,\sigma}$ the above symmetry
argument does not imply the vanishing of e.g.\ the matrix element 
$\langle {\rm ST}| N^{\rm (ps)}(0,0)|0\rangle$,
where $|{\rm ST}\rangle$ denotes a spin-triplet excitation
\cite{woyn:83}.  Given that symmetry does not force such matrix
elements to vanish we expect them to be nonzero, which then
immediately implies an algebraic decay like in (\ref{hfb}).

We emphasize, that to establish the exponential decay of $G_{nn}$ in
the above way it is essential that:
(1) there is an \emph{exact} symmetry in the
charge sector of the microscopic Hamiltonian (\ref{hamil}).
In general this symmetry may not be obvious (as is the case 
for the Hubbard model). 
(2) the
ground state is a singlet of the corresponding algebra.
Note that the above considerations still hold
in the presence of a magnetic field as the ground state 
(in the half-filled band) remains a singlet of the $\eta$-pairing SU(2).
(3) All charged (non-singlet) excitations are gapped.

We note that these conditions are fulfilled for the half-filled
Hubbard model on a bipartite lattice in \emph{any} dimension, provided
that $U$ is larger than the critical Mott-Hubbard value \cite{lieb:89}.

Analogous conclusions can be reached for spin-spin correlations in the
{\sl attractive} Hubbard model by employing discrete
symmetries of the Hamiltonian (\ref{hamil}). Under the particle-hole 
transformation for spin up
\begin{equation}
c_{j,\uparrow}\longrightarrow (-1)^j c^\dagger_{j,\uparrow}\ ,\quad
c_{j,\downarrow}\longrightarrow c_{j,\downarrow}
\end{equation}
the Hamiltonian (\ref{hamil}) transforms according to
${\cal H}(U)\longrightarrow {\cal H}(-U)$,
whereas $\eta$-pairing and spin SU(2) symmetries are interchanged
$S^a\leftrightarrow \eta^a$, $a=+,-,z$.
Furthermore the ground state of the attractive Hubbard model is a 
spin singlet.
This implies that spin-spin correlation functions in the attractive
Hubbard model decay exponentially at large distances
\begin{equation}
\langle S^z(t,x) S^z(0,0)\rangle\longrightarrow \exp(- \beta x)
\ ,\quad x\to\infty
\end{equation}
where $\beta >0$ for any $U<0$.

We are grateful to the Max-Planck-Institut f\"ur Physik komplexer
Systeme at Dresden for hospitality and to A.~O.\ Gogolin and 
A. Schadschneider for important discussions.
F.H.L.E.\ is supported by an EPSRC Advanced Fellowship.  H.F.\ is
supported in parts by the Deutsche For\-schungs\-gemeinschaft under
Grant No.\ Fr~737/2--3.
\end{multicols}

\end{document}